\documentclass[apj]{emulateapj}
\usepackage{natbib,url}
\usepackage{amssymb,amsmath}
\newcommand{\amm}{NH$_3$}
\newcommand{\dia}{N$_2$H$^+$}

\newcommand{\hd}{H$_2$D$^+$}

\newcommand{\kms}{km\,s$^{-1}$}
\newcommand{\cc}{cm$^{-3}$}

\shorttitle{}
\shortauthors{Friesen et al.}

\begin{document}

\title{Revealing \hd\, depletion and compact structure in starless and protostellar cores with ALMA}
\author{R. K. Friesen\altaffilmark{1}, J. Di Francesco\altaffilmark{2,3}, T. L. Bourke\altaffilmark{4}, P. Caselli\altaffilmark{5,6}, J. K. J{\o}rgensen\altaffilmark{7,8}, J. E. Pineda\altaffilmark{9}, M. Wong\altaffilmark{10}}
\altaffiltext{1}{Dunlap Institute for Astronomy and Astrophysics, University of Toronto, 50 St George St, Toronto, ON, M5S 3H4, Canada}
\email{friesen@dunlap.utoronto.ca}
\altaffiltext{2}{National Research Council Canada, Radio Astronomy Program, 5071 West Saanich Rd, Victoria, BC, V9E 2E7, Canada}
\altaffiltext{3}{Department of Physics and Astronomy, University of Victoria, Victoria, BC, V8P 1A1, Canada}
\altaffiltext{4}{Radio and Geoastronomy Division, Harvard Smithsonian Center for Astrophysics, MS-42, Cambridge, MA, 02138, USA}
\altaffiltext{5}{Max-Planck-Institute for Extraterrestrial Physics, Giessenbachstr. 1, D-85748 Garching, Germany}
\altaffiltext{6}{School of Physics and Astronomy, University of Leeds, Leeds LS2 9JT, UK}
\altaffiltext{7}{Niels Bohr Institute, University of Copenhagen, Juliane Maries Vej 30, 2100, Copenhagen \O ., Denmark}
\altaffiltext{8}{Centre for Star and Planet Formation, Natural History Museum of Denmark, University of Copenhagen, \O ster Voldgade 5-7, 1350, Copenhagen K., Denmark}
\altaffiltext{9}{Institute for Astronomy, ETH Z\"{u}rich, Wolfgang-Pauli-Strasse 27, CH-8093 Z\"{u}rich, Switzerland}
\altaffiltext{10}{University of British Columbia, Vancouver, BC, V6T 1Z4, Canada}
\begin{abstract}

We present Atacama Large Millimeter/submillimeter Array (ALMA) observations of the submillimeter dust continuum and \hd\, $1_{10}-1_{11}$ emission toward two evolved, potentially protostellar cores within the Ophiuchus molecular cloud, Oph A SM1 and SM1N. The data reveal small-scale condensations within both cores, with mass upper limits of $M \lesssim 0.02$~M$_\odot$ ($\sim 20$~M$_{Jup}$). The SM1 condensation is consistent with a nearly-symmetric Gaussian source with a width of only $37$~AU. The SM1N condensation is elongated, and extends 500~AU along its major axis. No evidence for substructure is seen in either source. A Jeans analysis indicates these sources are unlikely to fragment, suggesting that both will form single stars. \hd\, is only detected toward SM1N, offset from the continuum peak by $\sim 150-200$~AU. This offset may be due to either heating from an undetected, young, low luminosity protostellar source or first hydrostatic core, or HD (and consequently \hd) depletion in the cold centre of the condensation. We propose that SM1 is protostellar, and that the condensation detected by ALMA is a warm ($T \sim 30-50$~K) accretion disk. The less concentrated emission of the SM1N condensation suggests that it is still starless, but we cannot rule out the presence of a low-luminosity source, perhaps surrounded by a pseudodisk. These data reveal observationally the earliest stages of the formation of circumstellar accretion regions, and agree with theoretical predictions that disk formation can occur very early in the star formation process, coeval with or just after the formation of a first hydrostatic core or protostar. 

\end{abstract}

\keywords{ISM: molecules - stars: formation}

\section{Introduction}
\label{sec:intro}

Stars form within dense cores of molecular gas. Observations of the continuum emission from dust reveal that the density profiles of many starless cores can be well modelled with relatively flat centers that steepen to a power-law decrease at some radius. Dense cores that show high concentration - observationally, by comparing the integrated continuum flux to the central surface brightness - are more likely to contain embedded protostars \citep{walawender05}, including Very Low Luminosity Objects (VeLLOs) with internal luminosities $L_{int} \lesssim 0.1$~L$_\odot$ \citep{young04}. In analytic equilibrium models of dense cores, such as Bonnor-Ebert spheres, this behaviour is expected as there is a maximum concentration value for critically self-gravitating objects \citep{johnstone00} beyond which a core is unstable to collapse. As the central density of a collapsing core increases beyond $\sim 10^{11}$~\cc, the continuum emission becomes optically thick and the core begins to warm. The heated central molecular core achieves hydrostatic equilibrium, called the first hydrostatic core \citep[FHSC; ][]{larson69,masunaga98}, until H$_2$ is dissociated at $\sim 2000$~K (with the subsequent collapse forming the second hydrostatic core, or protostar). Concurrently, infalling material may form an accretion disk or pseudodisk with approximate size of a few AU to several hundred AU around a young protostar, due to the influence of magnetic fields \citep[e.g., ][]{galli_shu1,galli_shu2} and/or rotation \citep[e.g., ][]{bate14,commercon12}. The transition from prestellar core to FHSC or protostar, while clearly a vital step in the process of star formation, has not been well studied observationally due to the short lifetime of the FHSC stage, difficulties in establishing the relative evolutionary stage of objects, the small spatial scales involved, and the dearth of strong molecular tracers of cores at high densities. 

At the high densities ($n \gtrsim 10^6$\,\cc) and low temperatures ($T \le 20$\,K) characteristic of the interiors of highly evolved star-forming cores, many molecules deplete from the gas phase by freezing out onto dust grains \citep[see, e.g.,][]{bergin07}. Concurrently, the deuterium fraction of remaining gas-phase species increases dramatically above the cosmic abundance ratio \citep[$\mbox{[D/H]}  \sim 1.5 \times 10^{-5}$; ][]{oliveira03} due to increased production of \hd\, through the reaction of H$_3^+$ with HD in places (or regions) where CO is depleted \citep{roberts00}. At extreme densities, H$_3^+$ and its deuterated isotopologues may be the only observable species remaining in the gas phase \citep{walmsley04}. The lowest energy transitions of H$_3^+$ lie in the infrared and are unlikely to be excited in dense cores. Thus, \hd\, observations are crucial for our understanding of the chemistry (and magnetically regulated) dynamics just before a star is formed. The ortho-\hd\, $1_{10} - 1_{11}$ rotation transition at 372.4\,GHz is accessible from the ground but lies on the edge of a telluric water line and requires an exceptionally dry observing site. To date, \hd observations have only been successful with single-dish telescopes \citep[e.g., ][]{stark99, caselli03, caselli08, friesen10}. Based on the observational work of \citet{caselli03}, \citet{walmsley04} argued that the bright \hd\, emission toward L1544 could be explained by chemical models that include near-complete gas-phase depletion of species heavier than He within the central few thousand AU. In a further survey of 16 starless and protostellar cores, \citet{caselli08} found that strong \hd\, emission correlates with increasing core central density concentration, which itself correlates with growing instability to gravitational collapse.

\begin{figure}
\begin{center}
\includegraphics[width=0.5\textwidth,trim=0 0 150 50, clip]{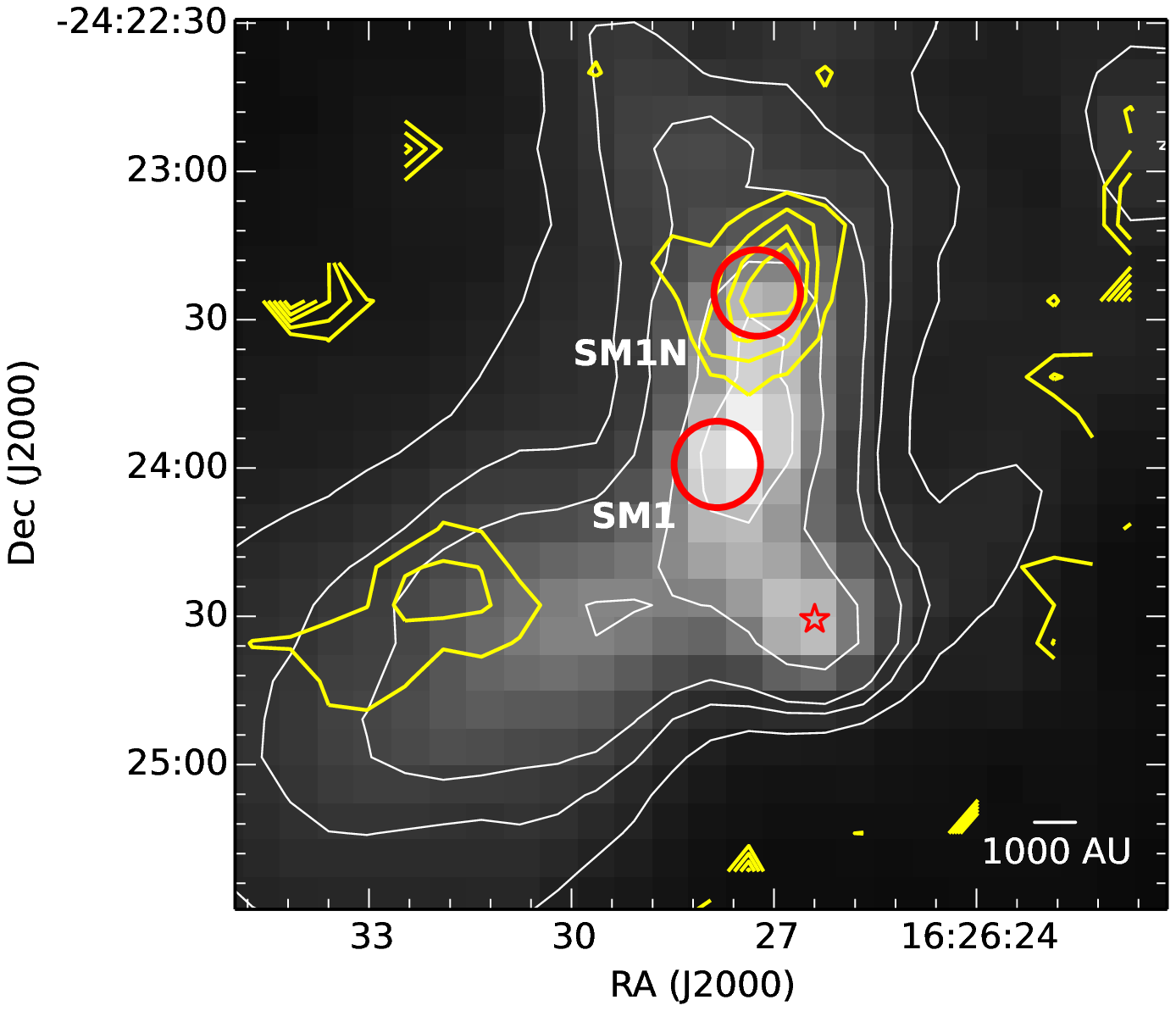}
\caption{Oph A in 850~\micron\, emission \citep[22.9\arcsec\, FWHM effective resolution, JCMT;][]{difran08}. The grey scale is in Jy~beam$^{-1}$. White contours show continuum contours at 0.5, 0.75, 1, 2, 3, 4~Jy~beam$^{-1}$. Yellow contours show \hd\, $1_{10}-1_{11}$ integrated intensity emission observed with the JCMT at 3, 6, 9 and 12 times the rms noise of 0.04~K~km~s$^{-1}$. Red circles show the  $\sim 17.5$\arcsec\, (FWHM) ALMA primary beam at 359~GHz for individual pointings toward SM1 and SM1N. The Class 0 protostar VLA 1623 is identified by the star. 
\label{fig:OphA}}
\end{center}
\end{figure}

\begin{deluxetable}{lcccc}
\tablecolumns{5}
\tablewidth{0pt}
\tablecaption{Source information and pointing phase centre \label{tab:targetInfo}}
\tablehead{
\colhead{Target} & 
\colhead{R.A.} & \colhead{Decl.} & \colhead{$S_{\tiny{peak}}$\tablenotemark{a}} & \colhead{$rms$\tablenotemark{a}} \\
\colhead{} & 
\colhead{J2000} & \colhead{J2000} & \colhead{mJy~beam$^{-1}$} & \colhead{mJy~beam$^{-1}$} 
}
\startdata
SM1    &  16:26:27.83 & -24:23:59.2 & 215 & 2 \\
SM1N &  16:26:27.25 & -24:23:24.6 & 3.0 & 0.3 
\enddata
\tablecomments{}
\tablenotetext{a}{Synthesized beam FWHM for SM1 and SM1N are given in the text.}
\end{deluxetable}

The Ophiuchus molecular cloud is the nearest example of ongoing, clustered star formation \citep[$d = 120$\,pc; ][]{loinard08}. The Oph A filament, shown in Figure \ref{fig:OphA} in 850~\micron\, continuum emission, is a 0.1~pc-length curved chain of $\sim 10$ clumps \citep{wilson99}, with high column densities $N(\mbox{H$_2$}) \sim 8 \times 10^{23}$\,cm$^{-2}$ and volume densities $n \sim 10^5 - 10^7$~\cc\, \citep{motte98,johnstone00}. Most of the cores are currently classified as starless, apart from the canonical Class 0 source VLA 1623, indicating Oph A is dynamically young. Apart from VLA 1623, the two brightest Oph A cores in millimeter continuum emission are SM1 and SM1N \citep{andre93}, which are separated by only $\sim 30$\arcsec\, (3600~AU), and have similar densities and masses \citep[$n \sim 10^6-10^8$~\cc\, and M $\sim 1-3~\mbox{M}_\odot$, respectively][]{motte98}. Neither core has been identified as protostellar through infrared observations. 

In this paper, we present Atacama Large Millimeter/submillimeter Array (ALMA) Cycle 0 submillimetre continuum and ortho-\hd\, 1$_{10}$ - 1$_{11}$ observations toward SM1 and SM1N at $\sim 50$~AU and $150$~AU spatial resolution (continuum and spectral line, respectively), as well as single-dish \hd\, observations made with the James Clerk Maxwell Telescope (JCMT). We describe the observations, data reduction, and imaging in \S\ref{sec:obs}. In \S\ref{sec:analysis}, we show that small-scale condensations are detected by ALMA within both SM1 and SM1N, and these condensations are resolved with effective radii of 37~AU and 400~AU, respectively. We derive masses of $\lesssim 0.02$~M$_\odot$ for both condensations, given reasonable assumptions of source temperature and dust opacity. We present the first interferometric detection of \hd, which we detect slightly offset from SM1N. In \S\ref{sec:disc}, we argue that the continuum emission detected by ALMA toward SM1 reveals a warm accretion disk surrounding an extremely young protostellar source. We further propose that the extended emission detected by ALMA toward SM1N is the first detection of a pseudodisk or circumstellar accretion region (CAR), possibly surrounding an extremely young protostellar source or first hydrostatic core (FHSC). 

\section{Observations and data reduction}
\label{sec:obs}

\subsection{ALMA}

Observations of the SM1 and SM1N cores were performed in 2012 August as part of the ALMA Early Science Cycle 0 program 2011.0.00396.S (PI Rachel Friesen) at 359~GHz and 372~GHz (ALMA Band 7). The core locations within Oph A are shown in Figure \ref{fig:OphA} (circles), while the pointing centres are listed in Table \ref{tab:targetInfo}. The data were obtained over two separate days with either 28 or 23 antennas available, but three antennas were later flagged in each dataset for having unusually high system temperatures. Baseline lengths ranged from 21~m to 402~m ($\sim 25 - 480$~k$\lambda$ at 359~GHz).  Four spectral windows were used, each with 234~MHz bandwidth and 3840 channels with widths of 61~kHz, or 0.05~\kms\, at 372~GHz. Two spectral windows were placed to observe the ortho-\hd\, 1$_{10}$--$1_{11}$ line at 372.213400~GHz and the \dia\, $J=$ 4--3 transition at 372.6725090~GHz \citep{pickett98}, while two additional spectral windows were used for continuum observations at 359.2~GHz and 359.4~GHz. Observations cycled between SM1, SM1N, and calibration sources in a given execution block. The ALMA primary beam at 359~GHz is $\sim 17.5$\arcsec, while the shortest baselines give a maximum angular size (the largest scale where structure is recoverable by the array) of $\sim 6$\arcsec. Since these observations were performed during Cycle 0, no compact array or total power measurements were available. 

Data were reduced and calibrated using the CASA 3.4 software package \citep{mcmullin07}. Flux calibration was performed using observations of Titan and Neptune, using the Butler-JPL-Horizons 2012 flux density models for solar system objects\footnotemark[1]\footnotetext[1]{ALMA Memo 594}. Bandpass calibration used observations of J1924-292 and J1517-243, and gain calibrations were performed through observations of J1625-254 and J1626-298. Calibration uncertainties were significantly lower for the spectral windows at 359~GHz than at 372~GHz due to the increased proximity to a strong telluric water line near 372~GHz. 

All four spectral windows were used in imaging the continuum data to maximize the total bandwidth. In the two spectral windows centred on the \hd\, and \dia\, emission lines, only line-free channels were used for continuum imaging. To image the SM1 condensation, we performed a multi-frequency synthesis deconvolution over all spectral windows and self-calibrated on the initial source model. The final image was created with uniform weighting of the $uv$-data to decrease side lobes and optimize angular resolution, giving a 0.45\arcsec $\times$ 0.34\arcsec\, beam full-width-half-maximum (FWHM) with a position angle of $-84$\arcdeg, and a 1 sigma noise sensitivity of 2.0~mJy~beam$^{-1}$. The lower peak flux density toward SM1N was not sufficiently bright to self-calibrate, and the final image was created with natural weighting of the $uv$-data to optimize the signal-to-noise ratio (SNR), using multi-scale, multi-frequency synthesis deconvolution to recover better the observed larger-scale structure of the source. This gave a beam FWHM of 0.60\arcsec $\times$ 0.41\arcsec\, with a position angle of $-75$\arcdeg, and a 1 sigma noise sensitivity of 0.3~mJy~beam$^{-1}$. 

Before imaging the line spectral windows, continuum emission in each channel was removed with a linear fit. To increase the surface brightness sensitivity, the \hd\, data were tapered in the $uv$-plane to a final angular resolution of 1\arcsec.3, or $\sim 150$~AU, while the velocity resolution was decreased to 0.15~\kms\, velocity channels.  The final rms noise in the tapered data is 0.07~Jy~beam$^{-1}$ (0.36~K in the scale of brightness temperature $T_b$) per 0.15~\kms\, velocity channel. 

\subsection{JCMT}

Ortho-H$_{2}$D$^{+}$ 1$_{10}$--1$_{11}$ and N$_{2}$H$^{+}$ 4--3 emission from Oph A was observed with the JCMT using the 16-element Heterodyne Array Receiver Programme (HARP) focal-plane array \citep{smith08} and the AutoCorrelation Spectrometer Imaging System (ACSIS) back-end \citep{hovey00}. HARP consists of 16 receptors arranged in a 4$\times$4 square pattern with an on-sky spacing between receptors of 30\arcsec.  Data were obtained by pointing HARP at one of three target positions along Oph A, and observing in a ``jiggle" mode to obtain three adjacent fully sampled maps, each $\sim 2$\arcmin\, in diameter due to field rotation.  Sky subtraction was accomplished by position switching between the target positions and locations 900\arcsec\, separated in elevation.  HARP was tuned to 372.5 GHz and ACSIS configured to provide two windows, centred on the \hd\, and \dia\, lines, each with 250 MHz band width and 4096 channels. Each channel was 61~kHz wide (0.049~\kms\, at 372~GHz).  The velocity resolution of ACSIS data is a factor of $\sim 1.2$ the configured channel spacing, or in this case 0.059~\kms.

Observing proceeded over three semesters, under project IDs M09AC08, M09BAC14, and  M10AC03.  All observations were made during very dry weather conditions, i.e.,  $\tau_{225} < 0.05$, to ensure maximum sensitivity to the ortho-H$_{2}$D$^{+}$  1$_{10}$--1$_{11}$ and the N$_{2}$H$^{+}$ 4--3 lines.   Standard focus calibration observations were performed on nearby bright objects like R Aql, and pointing calibrations observations were conducted on objects such as CRL618, both before the observations start and every $\sim 60-90$ minutes.  Flux calibration was monitored by observing when possible spectral line standards like W75N at a variety of frequencies.  The aperture efficiency of HARP at 372.5 GHz is not well characterized, and is presumed to be $\sim 0.63$.

The data were reduced consistently using common routines and procedures in the STARLINK reduction package.  Each integration was visually checked for 
baseline ripples, absent detectors or extremely large spikes and specific integrations where these occurred were removed from the data ensemble.  In particular, receptor H03 was unavailable during most of the data; at various times no more than two other receptors were also unavailable.  Other spikes were identified and removed using a standard methodology created by developers at the Joint Astronomy Centre.  Spectral baselines were subtracted, frequency axes converted to velocities, and spectra trimmed using STARLINK scripts kindly provided by T. van Kempen.  All remaining data were co-added into pixels 7.5\arcsec\, $\times$ 7.5\arcsec\, in size, placed on the WCS framework, and written out as FITS format files using other standard STARLINK routines.  

\section{Results and Analysis}
\label{sec:analysis}

\begin{figure*}
\includegraphics*[trim=0 0 0 240,clip,width=7in]{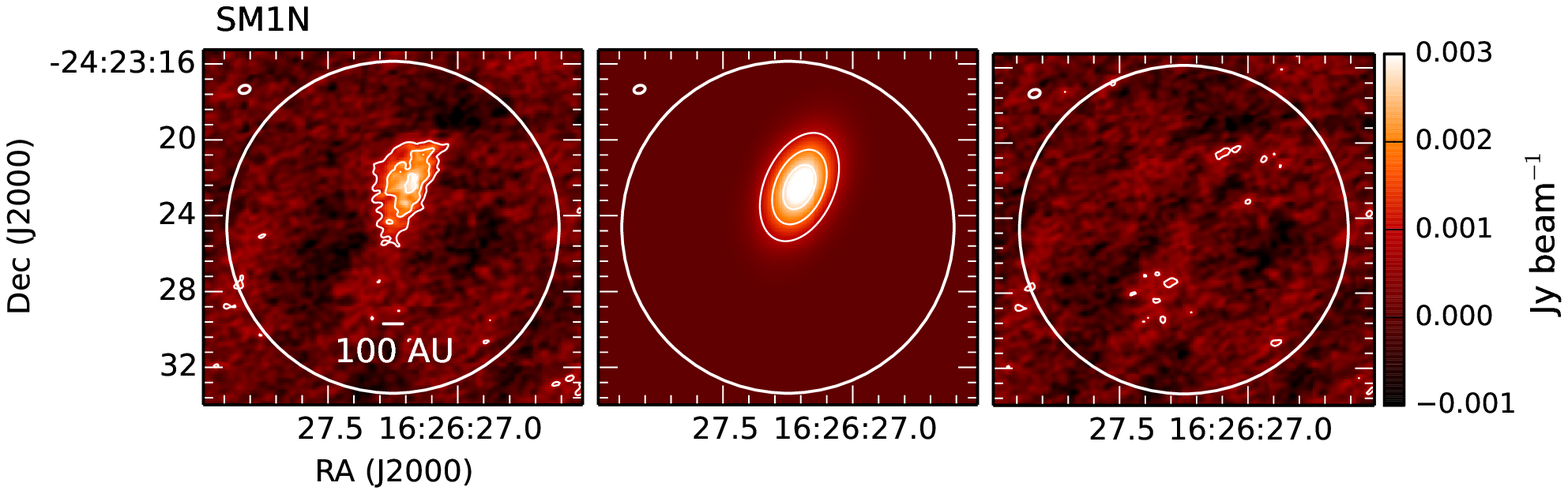}
\includegraphics*[trim=0 0 0 240,clip,width=7in]{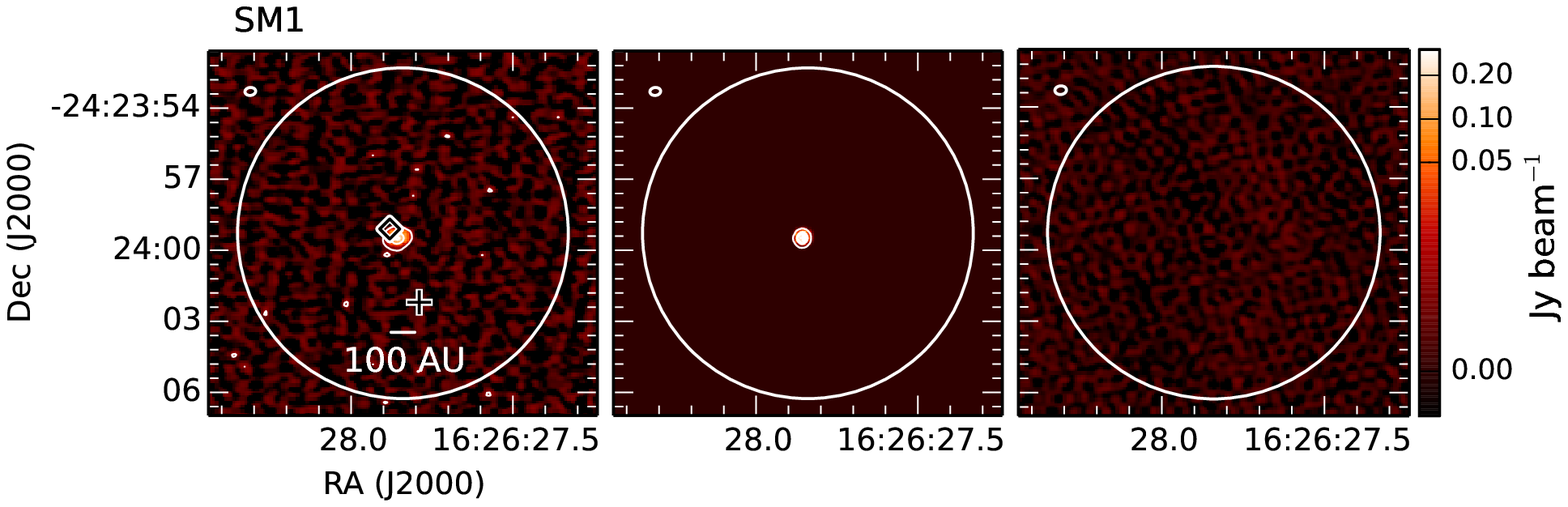}
\caption{Continuum image at 359~GHz (left), image of the best-fit Gaussian model to the $uv$-data (middle), and model visibility-subtracted image (right) for SM1N (top) and SM1 (bottom).  ALMA synthesized beams are shown in the top left corner for each source, while the large white circle shows the $\sim 17.5$\arcsec\, (FWHM) ALMA primary beam at 359~GHz.  For both sources, the colour scale is in Jy~beam$^{-1}$, but note that SM1 is shown with log scale in intensity. For SM1, white contours show 6~mJy~beam$^{-1}$, 100~mJy~beam$^{-1}$, and 200~mJy~beam$^{-1}$ (where 6~mJy~beam$^{-1}$ is 3 sigma); for SM1N, white contours show 0.3~mJy~beam$^{-1}$, 0.9~mJy~beam$^{-1}$, and 1.2~mJy~beam$^{-1}$ (where 0.3~mJy~beam$^{-1}$ is 3 sigma).  Toward SM1, the cross shows the location of a 6~cm radio emission detection, while the diamond shows a faint X-ray source \citep{gagne04}. \label{fig:modImage}}
\end{figure*}

\begin{deluxetable*}{lcccccc}
\tablecolumns{7}
\tablewidth{0pt}
\tablecaption{Results from $uv$-modeling of continuum data \label{tab:modelResults}}
\tablehead{
\colhead{Target} & 
\colhead{R. A.} & \colhead{Decl.} & \colhead{$S_\nu$} & \colhead{Major axis} & \colhead{Minor-major axis ratio} & \colhead{Position angle}  \\
\colhead{} & \colhead{} & \colhead{} & \colhead{Jy} & \colhead{\arcsec} &  \colhead{} & \colhead{\arcdeg}
}
\startdata
SM1    &  16:26:27.86 &  -24:23:59.5 & 0.317(1) & 0.314(1) & 0.908(7) & -6(2) \\
SM1N &  16:26:27.20 &  -24:23:22.4 & 0.15(1) & 4.2(2) & 0.61(3) & -24(3) \\ 
\enddata
\tablecomments{}
\end{deluxetable*}

With ALMA, we detected small-scale structure toward both SM1 and SM1N in the continuum. Here, we call the ALMA-detected structures `condensations', and retain the `core' nomenclature for the structures identified through single-dish observations. The ALMA continuum images are shown in Figure \ref{fig:modImage}, and Table \ref{tab:targetInfo} lists the peak flux density and rms noise observed by ALMA for both sources. Toward SM1, we find a small but resolved source with a peak flux density of 215~mJy~beam$^{-1}$ and total flux density of 321~mJy. The emission peak in the interferometer data is offset by $\sim 4-6$\arcsec\, from the SM1 core location identified in dust continuum single-dish observations \citep{johnstone00,motte98}. The peak flux density observed by ALMA is only $\sim 5$~\% of the 850~\micron\, single-dish peak flux density, while the total flux density is only $\sim 1$~\% of the single-dish flux density. Toward SM1N, we find a fainter, larger, and elongated source with a peak flux density of 3~mJy~beam$^{-1}$ and a total flux density of 150~mJy. The condensation is offset 6\arcsec\, from the core location as identified through 1.3~mm single-dish emission (\citeauthor{motte98}). The observed ALMA peak flux density is only $\sim 0.1$~\% of the single-dish flux density, where the 1.3~mm flux was determined after subtraction of the larger-scale Oph A clump emission, and further scaled to 359~GHz assuming typical starless core dust emissivity (see \S \ref{sec:masses} for more discussion of dust emissivity). We do not have a single-dish total flux measurement for SM1N, but compare the core and condensation masses further in \S \ref{sec:masses}. 

\begin{figure}
\includegraphics[width=0.5\textwidth, trim=17 0 0 0, clip]{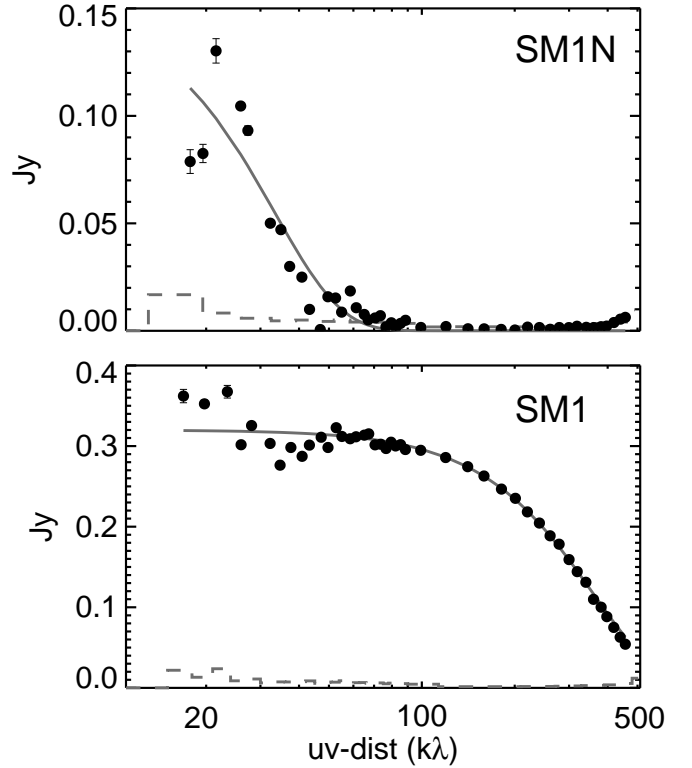}
\caption{Black points show circularly averaged visibilities for SM1N (top) and SM1 (bottom). Error bars show the standard error. The solid grey line shows the best-fitting Gaussian model to the $uv$-visibilities, while the dashed grey line shows the expected value for zero flux. \label{fig:modUV}}
\end{figure}

Figure \ref{fig:modUV} (black points) shows the circularly averaged $uv$-data for the SM1 and SM1N condensations. We show the continuum data in the 359~GHz spectral windows only, as the noise was greater in the 372~GHz windows due to their proximity to the telluric water lines. For each source, we shifted the phase centre of the $uv$-data to match the condensation peak flux position before performing circular averaging. The grey lines show the best-fit Gaussian model to the visibilities, described further in the next section. The amplitude expectations for zero flux (grey dashed lines) were calculated using the \emph{uvamp} task in the Miriad software package \citep{sault95}. 

The $uv$-data further emphasize what can be seen in the continuum images for each source. The SM1 condensation is resolved, with significant flux above the zero expectation value to the longest baselines (smallest angular scales) we measure. Furthermore, there is a break in the visibility amplitudes at $uv$-distances $\lesssim 60$~k$\lambda$, where the amplitudes oscillate around the apparent flux density limit. This beating pattern indicates that there is additional emission on larger angular scales (as measured at small $uv$ distances), but it is not connected to the SM1 condensation itself. The SM1 condensation is thus distinct from the larger-scale core, and is not an extrapolation of the emission from larger scales. In contrast to SM1, the $uv$-data for SM1N show that the SM1N condensation is extended, and has no evidence for additional compact structure at $uv$-distances $\gtrsim 60$~k$\lambda$ at flux densities $\gtrsim 2-4$~mJy. 

We additionally present here the first interferometric detection of \hd\, toward a starless or protostellar core. The single-dish \hd\, data, shown in Figure \ref{fig:OphA} (yellow contours), reveal two distinct \hd\, emission peaks within Oph A. \hd\, emission is found toward SM1N, and also near Oph AN6, a quiescent, starless core identified in \dia\, 1-0 emission \citep{difran04}. With ALMA, we detect \hd\, emission toward SM1N, but not SM1, with a peak line brightness of 0.25~Jy~beam$^{-1}$ (1.3~K in $T_b$ scale).  Figure \ref{fig:hdspec} (top) shows the \hd\, integrated intensity toward SM1N. Comparison of the \hd\, integrated intensity peak (black contours begin at 3 sigma) with the continuum contours (white) shows that the emission peaks are offset by $\sim 1.5$\arcsec\, ($\sim 150-200$~AU at 120~pc). Figure \ref{fig:hdspec} additionally shows the \hd\, spectra, taken at the peak of the ALMA integrated intensity map, from the JCMT and ALMA observations. The line velocities of peak brightness are similar between the single-dish and the interferometric data.  

\dia\, 4-3 emission was also detected toward both sources, but is heavily self-absorbed or resolved out by the interferometer, and we do not show it here. Additional observations that include shorter baselines or total power information are needed to understand the structure of the \dia\, emission. We note, however, that the \dia\, emission does not peak toward the continuum structures in either source, either in the integrated intensity or channel maps. We do not discuss the \dia\, data further in this paper. 

\subsection{Continuum $uv$-modeling}

Given the relatively simple structure seen in the continuum images, we fit both the SM1 and SM1N condensations with single 2D Gaussians in the $uv$-plane using the CASA task \emph{uvmodelfit}. This task performs a least-squares minimization of a model to the observed visibilities in amplitude and phase. By fitting directly source models to the observed visibilities rather than to the imaged data, we circumvent the deconvolution step and analyze the data in the domain where the measurements are made. For both sources, only the spectral windows at 359~GHz were used in this analysis. We then subtracted the best-fit model visibilities from the observed visibilities, and imaged the residuals to determine the quality of the fit. Curves representing the best-fit 2D Gaussian model radially-averaged visibility amplitude as a function of \emph{uv}-distance for each source are overlaid on the data points in Figure \ref{fig:modUV}. The fit results are listed in Table \ref{tab:modelResults}. Figure \ref{fig:modImage} shows the ALMA continuum images, the best-fit model images, and the residual images for both SM1 and SM1N.

The SM1 condensation is well fit by a single 2D Gaussian component with a major axis, $a_{maj} =  0.31$\arcsec\, ($\sim 37$~AU at 120~pc), a minor-major axis ratio of 0.91, position angle (PA) of $-6$\arcdeg, and total flux $I = 0.317$~Jy. The residual in the model-subtracted image, made with uniform weighting of the visibilities, is $1$~mJy, a factor of two less than the 1 sigma rms of the original image. A faint striping pattern of positive and negative residuals is present, however (see Figure \ref{fig:modImage}). As described in the previous section, the circularly averaged plot of the visibility amplitudes of SM1 shows evidence of ringing at short $uv$-distances, indicating that additional structure is present at larger scales than the bright, 0.31\arcsec\, (Gaussian width) emission peak. Emission is seen to the west of the SM1 condensation with peak flux densities of $\sim 2$~mJy~beam$^{-1}$. The offset from the SM1 condensation, and the pattern of negative and positive striping suggests that the emission is coming from unresolved structure such as, for example, the western edge of the Oph A filament, rather than an underlying structure similar in size and brightness to the SM1N condensation (discussed further below). Further observations with additional short-spacing information are needed to discern the source of the larger-scale emission. 

SM1N is also moderately well-fit by a single 2D Gaussian component, but with significantly greater extent, larger aspect ratio, and lower flux than SM1. The emission peak is shifted by 2.2\arcsec\, from the pointing phase centre, but still falls well within the 17.5\arcsec\, (FWHM) ALMA primary beam. The best-fit Gaussian gives $a_{maj} = 4.2$\arcsec\, ($\sim 520$~AU at 120~pc),  a minor-major axis ratio of 0.61, PA of $-24$\arcdeg, and total flux $S_\nu = 0.15$~Jy. Similar to SM1, there may be additional flux at the shortest $uv$-distances in the circularly averaged plot of the visibility amplitudes, but our data do not have the sensitivity and short-spacing information to determine its structure. As previously described, the visibility amplitudes at large \emph{uv}-distances in Figure \ref{fig:modUV} are consistent with the expectation for zero flux, ruling out any additional, unresolved structure at flux density $\gtrsim 2-4$~mJy. 

\subsection{Masses of the condensations}
\label{sec:masses}

For both sources, the total flux density, $S_\nu$, derived from the models agrees with the total flux density summed over the continuum maps. Assuming the dust emission is optically thin, we directly calculate the mass from $S_\nu$ for each source, following

\begin{equation}
M = \frac{d^2 S_\nu}{\kappa_\nu B_\nu(T_d)},
\end{equation}

\noindent where $d = 120$~pc is the distance to Oph A \citep{loinard08}. The dust opacity per unit mass at frequency $\nu$, $\kappa_\nu$, is parameterized as $\kappa_\nu = \kappa_0 \nu^\beta$, where $\kappa_0$ is the dust opacity per unit mass normalized at some frequency $\nu_0$, and $\beta$ is the opacity spectral index. $B_\nu (T_d)$ is the Planck function at frequency $\nu$ and dust temperature $T_d$. 

We constrain $T_d$ by assuming it is similar to the gas temperature, given the expected coupling between gas and dust at the densities of the cores. We then determine the kinetic gas temperature through simultaneous hyperfine fitting of \amm\, (1,1) and (2,2) line emission following \citet{friesen09}, observed over the entire Oph A core with the Australia Telescope Compact Array (10\arcsec.6 $\times$ 7\arcsec.0 FWHM, or $\sim 1000$~AU; Friesen et al., in preparation). We find kinetic gas temperatures $T_K = 16.0 \pm 0.7$~K toward SM1N. SM1 lies at the edge of the Oph A filament as seen in \amm\, emission, with consequently lower SNR. The line strengths of the main components of the \amm\, (1,1) and (2,2) lines are similar at the SM1 location, however, indicating that the gas temperature is at least $\sim 30$~K. Given the SNR and the similarity of the line strengths, we cannot constrain the temperature more accurately with this thermometer. We also note that the spatial scales of the observations from which the temperatures are derived are greater than what ALMA can probe at these wavelengths. The true gas and dust temperature at $\sim 50-400$~AU scales may therefore be lower if the cores are starless, or greater if the cores are protostellar. 

The dust opacity, $\kappa_\nu$, and opacity spectral index, $\beta$, are additional sources of uncertainty. Typical dust opacities in the interstellar medium generally assume $\beta \sim 1.8$, in agreement with models of ice-coated dust grains \citep[OH5; ][]{ossen94}. Grain growth can decrease the spectral index. In protostellar disks, multiple studies have shown that dust coagulation decreases the observed index  to $\beta \sim 1$ or lower \citep[e.g., ][]{banzatti13}. In the limit that the dust emission is optically thick, we expect to see a pure blackbody source at temperature $T_d$ and $\beta = 0$. To estimate the mass of the SM1 and SM1N condensations, we normalize $\kappa_0 = 0.1$~cm$^2$~g$^{-1}$ at 1000~GHz \citep{hildebrand83}. Setting $\beta = 1.8$ gives $\kappa_\nu = 0.016$~cm$^2$~g$^{-1}$ at 359~GHz ($\sim 835$~\micron), in agreement with the OH5 model, while $\kappa_\nu = 0.036$~cm$^2$~g$^{-1}$ with $\beta = 1$. 

For SM1, we can further constrain the spectral index of the dust opacity through Submillimeter Array (SMA) observations at 227~GHz (1.3~mm; Bourke et al.  in prep). We use data only on baselines $> 20$~k$\lambda$ (giving a synthesized beam of $\sim 3$\arcsec), to exclude emission resolved out by the ALMA observations, and find a point source flux of $133 \pm 3$~mJy toward SM1. Assuming the emission is in the Rayleigh-Jeans regime at both 227~GHz and 359~GHz \citep[following, e.g.,][]{friesen05}, and $\sim 10$~\% flux calibration uncertainty, we find $\beta = 0.4 \pm 0.3$. This low value of $\beta$ is suggestive of grain growth in SM1. If we have overestimated the 227~GHz flux by including emission resolved out by ALMA, however, the $\beta$ thus derived will be a lower limit. While similarly low values of $\beta$ have been reported toward young disks \citep{tobin13}, more frequently $\beta = 1$ is assumed to fit the dust emission in disks around protostars. Assuming a minimum temperature for the SM1 condensation of $\sim 30$~K and $\beta = 0.4$, we find $M = 0.004$~M$_\odot$.  If, instead, SM1 is a warm accretion disk, for example, we may expect $\beta = 1$ and $T_d \sim 150$~K \citep[e.g., ][]{zapata13}, giving a disk mass $M = 0.001$~M$_\odot$. Archival Herschel data, however, described further in \S\ref{sec:evol}, show no point-source emission toward SM1. At a uniform temperature $T \gtrsim 50$~K, SM1 would be sufficiently bright at 70~\micron\, to be detected in the archival Herschel data, and we can thus rule out such high temperatures averaged over the condensation. A circumstellar disk heated by a central protostar will have a temperature gradient, with warmer temperatures at small radii that would remain consistent with our data. 

Given the mass, we further calculate the mean volume density, $<n>$, assuming a spherical shape with an effective radius $R_{eff} = \sqrt{a_{maj}~a_{min}}$. Regardless of the choice of $\beta$ or $T_d$, the SM1 condensation has a remarkably high mean density. Setting $T_d = 30$~K, $R_{eff} = 37$~AU, and $\beta = 0.4$, we find $<n> = 3 \times 10^{9}$~\cc. The Jeans length for these parameters, $\ell_J \simeq \sqrt{c_s / G \rho} \sim 80$~AU, is slightly greater than the observed source size. At temperatures $\lesssim 50$~K, however, the free-fall timescale associated with SM1's derived mean density $t_{ff} \sim 10^3$~yr, making it highly unlikely to be observed in this state. We conclude that the SM1 condensation is most likely a warm disk with an average temperature $T \sim 30-50$~K and $M \sim 0.002 - 0.02$~M$_\odot$ ($\sim 2 - 20$~M$_{Jup}$), where the lowest values correspond to a combination of high $T$ and low $\beta$. For a uniform disk with a radius $R_{eff}$, $\beta = 0.4$ and $T_d \sim 30$~K, the corresponding surface density $\Sigma = 8$~g~cm$^{-2}$ (including both gas and dust).  

\begin{figure}
\includegraphics[width=0.5\textwidth,trim=0 0 360 15, clip]{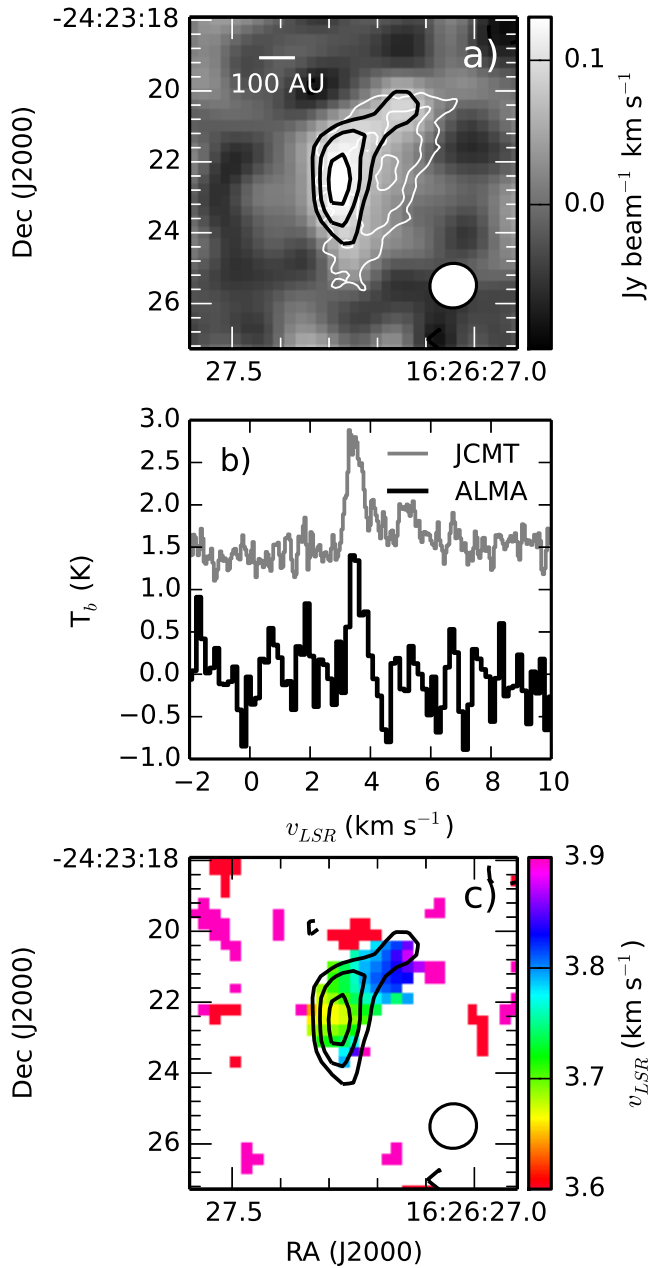}
\caption{Top: Integrated \hd\, $1_{10}-1_{11}$ intensity toward the SM1N core (greyscale) at 1\arcsec.3 (FWHM) resolution, with black contours at $3~\sigma$, $4~\sigma$, and $5~\sigma$, where the 1 sigma rms noise is 0.025~Jy~beam$^{-1}$~km~s$^{-1}$. White contours show the continuum emission as in Figure \ref{fig:OphA}. Middle: ALMA (black) and JCMT (grey) \hd\, spectra toward SM1N. Both spectra are taken at the \hd\, integrated intensity peak in the ALMA data. For better comparison, the ALMA data have been converted to brightness temperature units, $T_b$, while the JCMT data have been converted to $T_{mb}$ units assuming a main beam efficiency $\eta_{mb} = 0.63$. Note that the JCMT beam FWHM is $\sim 13$\arcsec. Bottom: \hd\, $v_{LSR}$ as determined through Gaussian fitting of the line profile.  \label{fig:hdspec}}
\end{figure}

For SM1N, we expect the \amm-derived gas temperature, $T_{gas} = 16$~K, to be a good first approximation to the true dust temperature. If we assume the typical dense core $\beta = 1.8$, we find the SM1N condensation has a mass of 0.02~M$_\odot$. For a range of temperatures between 10~K and 25~K, the mass ranges from 0.05~M$_\odot$ to 0.01~M$_\odot$. Lowering $\beta$ to 1 decreases the derived mass by a factor $\sim 2$ in this temperature range.  With $R_{eff} = 390$~AU, SM1N's mean density is large ($<n> \sim 10^7$~\cc), but is similar to that derived by \citet[][; $n = 3.6 \times 10^7$~\cc]{motte98} for SM1N, using a multiresolution wavelet analysis to separate small- and large-scale structures in a 1.3~mm single-dish continuum map (although their source size is significantly larger). For comparison, the combination of mass and effective radius reported by \citet{johnstone00} through 850~\micron\, observations at 15\arcsec\, resolution gives a mean density for SM1N of $n \sim 2 \times 10^5$~\cc.  Given a density $n = 10^7$~\cc, the free-fall time $t_{ff} \sim 4 \times 10^4$~yr.  For SM1N we find $\ell_J \sim 1000$~AU for a temperature of 15~K, greater than SM1N's effective radius. 

While the uncertainty in the derived masses is thus uncertain to factors of a few, we find that the condensations associated with both SM1 and SM1N have masses $M \lesssim 0.02$~M$_\odot$. For comparison, the brown dwarf mass limit $M_{bd} \sim 0.01$~M$_\odot$ (13~M$_{Jup}$) is very similar to the masses we determine for the ALMA-detected condensations. These masses are also $\gtrsim 100$ times smaller than those derived from single-dish observations \citep[1-3~M$_\odot$; ][]{motte98,johnstone00}. If either SM1 or SM1N are protostellar, they are clearly early stage objects, where the mass in any small-scale structure or disk is substantially less than that remaining in the envelope. 

\subsection{Ortho-\hd\, line analysis}

We fit the \hd\, emission profile detected by ALMA with a single 1D Gaussian at pixels where the peak line emission is greater than three times the rms noise.  Within the $3~\sigma$ contours of the integrated intensity map, we find a mean line-of-sight velocity relative to the local standard of rest, $v_{LSR} = 3.75$~\kms, and a mean line width (FWHM) $\Delta v = 0.46$~\kms. Typical uncertainties on these values are $\sim 0.06$~\kms\, and $\sim 0.1$~\kms, respectively. Figure \ref{fig:hdspec} (bottom) shows the spatial distribution of  $v_{LSR}$ obtained from \hd. A small gradient in $v_{LSR}$,  $\sim 0.12$~\kms\, over $\sim 200$~AU, can be seen along the long axis of SM1N, with line-of-sight velocity increasing to the northwest, but we are unable to perform a robust fit to the gradient given the low SNR of the \hd\, emission. A gradient in $v_{LSR}$ is also seen along the long axis of SM1N in the \amm\, (1,1) data, but with the line-of-sight velocity increasing to the southeast, i.e. opposite the direction seen in the \hd\, emission. The \amm\, gradient is in agreement with the larger-scale trend seen in \dia\, emission \citep{difran04}. A similar direction reversal of the velocity gradient at small scales was seen in L1544 \citep{caselli02,crapsi07}. This intriguing potential reversal in velocity gradients can be probed with greater sensitivity line observations of the small-scale SM1N structure. 

Since \hd\, has a low molecular weight ($m_{\mbox{\tiny{\hd}}} = 4.01~m_{\mbox{\tiny{H}}}$), the thermal velocity dispersion, $\sigma_t = (k_B T / m_{\mbox{{\tiny{\hd}}}})^{0.5}$, for a given gas temperature, $T$, is relatively large. If purely due to thermal motions, the mean \hd\, line width (FWHM) $\Delta v = 0.46$~\kms\, observed by ALMA toward SM1N corresponds to a temperature of 18.5~K, similar to the temperature we assume for SM1N based on the \amm\, analysis. In a few pixels, the line widths decrease to $\lesssim 0.4$~\kms, suggesting a lower temperature limit $\lesssim 14$~K. The $\sim 0.1$~\kms\, uncertainty in the fits, however, limits any temperature constraints to approximately a few K. The line widths observed by the JCMT are slightly larger, $\Delta v = 0.63$~\kms\, toward SM1N (at 13\arcsec\, FWHM resolution), but correspond to a gas temperature of 35~K if purely thermally driven. If we assume instead a gas temperature of 16~K, we find a non-thermal velocity dispersion $\sigma_{\mbox{\tiny{NT}}} = \sqrt{\sigma_v - \sigma_t} = 0.2$~\kms\, for the \hd\, emission at 13\arcsec\, resolution, slightly less than the sound speed at 16~K. The non-thermal motions traced by \hd\, in Oph A are thus subsonic on $\sim 1500$~AU scales (13\arcsec\, at 120~pc), and decline to extremely low values on $\sim 140$~AU scales. This suggests that unless the SM1N condensation is significantly colder than 16~K, the \hd\, emission is not tracing any significant non-thermal motions due to, e.g., infall. Even if the gas temperature declines to $\sim 5-10$~K on the small scales traced by ALMA, any non-thermal motions will remain subsonic, on average.

We next determine the ortho-\hd\, column density, $N(\mbox{ortho-\hd})$, following \citet{friesen10}. We first derive an opacity $\tau$ assuming an excitation temperature $T_{ex}$, 

\begin{equation}
\tau = -\mbox{ln} \bigg( 1 - \frac{T_{mb}}{J(T_{ex}) - J(T_{bg})}\bigg) ,
\end{equation}

\noindent where $T_{mb}$ is the peak main-beam temperature of the \hd\, emission line, and $J(T_{ex})$ and $J(T_{bg})$ are the equivalent Rayleigh-Jeans excitation and background temperatures, respectively. We then calculate the column density \citep{vastel06}, 

\begin{align}
N(\mbox{ortho-\hd}) =& ~\frac{8\pi}{\lambda^3A_{ul}}\frac{Q_{rot}(T_{ex})}{g_u}  \\ \nonumber
					&\times \frac{\exp(E_u/kT_{ex})}{\exp(h\nu/kT_{ex})-1} \int \tau \mbox{d}v .
\end{align}

Here, the statistical weight $g_{u} = 9$, $E_u/k = 17.8$~K, and $A_{ul} = 1.08 \times 10^{-4}$~s$^{-1}$ for the ortho-\hd\, $1_{11} - 1_{10}$ transition \citep{raman04}. The integral $\int \tau \mbox{d}v = \frac{1}{2}\sqrt{\pi/(\ln(2))} \tau \Delta v$. We estimate the partition function $Q_{rot}$ by reducing the \hd\, level structure to a 2-level system following \citet{caselli08}. Since the energy of the first excited state above ground is $E/k = 17.8$\,K and that of the second excited state is $E/k = 110$\,K, we expect this approximation to be valid in SM1N. The partition function depends on $T_{ex}$. \citeauthor{caselli08} found that $T_{ex}$ in \hd\, is similar to, or slightly less than the kinetic gas temperature toward several starless and protostellar cores. For SM1N, we thus set $T_{ex} = 12$\,K, less than the gas temperature $T_K = 16$~K.  This choice also allows best comparisons with single-dish \hd\, results in Ophiuchus \citep[e.g.,][]{friesen10}. For similar physical conditions, \citeauthor{friesen10} showed that the derived ortho-\hd\, column density will underestimate the true value by $\sim 30$~\% if $T_{ex} = 10$~K, and overestimate the true value by $\sim 20$~\% if $T_{ex} = 14$~K.  

We find a maximum ortho-\hd\, column density $N(\mbox{ortho-\hd}) = 1.2 \times 10^{13}$~cm$^{-2}$ at the integrated intensity \hd\, peak. Based on the rms noise in the data and the mean \hd\, line width (where detected), we estimate the $1~\sigma$ uncertainty on the column density, $\sigma_N = 2.5 \times 10^{12}$~cm$^{-2}$. Note that this uncertainty does not take into account any systematic calibration uncertainties or the uncertainty in $T_{ex}$. Since \hd\, is not detected toward the SM1N continuum peak, we therefore find a $3~\sigma$ upper limit on the ortho-\hd\, column density of $7.5 \times 10^{12}$~cm$^{-2}$ at that location. 

We derive the line-of-sight abundance of ortho-\hd\, across SM1N by first imaging the continuum data with the same taper. We then calculate the H$_2$ column density following $N(\mbox{H}_2) = S_\nu / (\Omega_b \mu m_{\mbox{\tiny{H}}} B(T_d))$, where $\Omega_b$ is the beam solid angle, $\mu = 2.8$ is the mean molecular weight of interstellar material in a molecular cloud per hydrogen molecule, and we set $T_d = 17$~K in agreement with the \amm-derived kinetic gas temperature. The ortho-\hd\, abundance is then $X(\mbox{ortho-\hd}) = N(\mbox{ortho-\hd})/N(\mbox{H}_2)$. The maximum abundance thus derived is $X(\mbox{ortho-\hd}) = 10^{-10}$, with a 3 sigma upper limit at the continuum peak of $X(\mbox{ortho-\hd}) \leq 5 \times 10^{-11}$. 

For comparison, a similar analysis of submillimetre continuum and \hd\, observations at 15\arcsec\, resolution in Oph A gives $N(\mbox{ortho-\hd}) = 1.4 \times 10^{13}$~cm$^{-2}$, H$_2$ column density $N(\mbox{H}_2) = 1.5 \times 10^{23}$~cm$^{-2}$, and an ortho-\hd\, abundance $X(\mbox{ortho-\hd}) = 9 \times 10^{-11}$ toward SM1N. The maximum ortho-\hd\, abundance observed by ALMA is thus similar to that observed at larger scales, but ALMA reveals a drop in abundance toward the continuum peak by a factor $\gtrsim 2$. 

Given the warmer temperature estimated for the SM1 disk, the lack of \hd\, detection is likely the result of the processes that form \hd\, in abundance requiring both cold temperatures and depleted CO. At temperatures $\gtrsim 20$~K, CO is able to evaporate from dust grains back into the gas phase and interact with H$_3^+$ and \hd, lowering the \hd\, abundance \citep{caselli08}. 

\section{Discussion}
\label{sec:disc}

\subsection{The evolutionary states of SM1 and SM1N}
\label{sec:evol}

Embedded protostellar sources are most readily identified as point sources in the infrared, for example through Spitzer or Herschel imaging at 24~\micron\, or 70~\micron\, wavelengths. \citet{dunham08} find a strong correlation between protostellar internal luminosity, $L_{int}$, and the 70~\micron\, flux of an embedded YSO (their Equation 2). We use this relation to place limits on any embedded sources within SM1 and SM1N, using archival 70~\micron\, data from the PACS instrument \citep{poglitsch} on the Herschel Space Observatory\footnotemark[2]\footnotetext[2]{Data are from program OT1\_pabraham\_3. Herschel is an ESA space observatory with science instruments provided by European-led Principal Investigator consortia and with important participation from NASA.}. Toward Oph A, the only point source visible at 70~\micron\, is the Class 0 protostar VLA 1623. Since the total flux density of a point source is equal to the flux density within a telescope beam, we can calculate the upper limit on $L_{int}$ using the $3~\sigma$ noise value at 70~\micron\, within a radius of 5.2\arcsec\, (the Herschel beam FWHM at 70~\micron).  An arc of bright emission is present along the eastern edge of Oph A, however. Both SM1 and SM1N lie near, or along, this ridge of emission, and the upper limits we derive are consequently relatively high. After applying a colour correction \citep{pezzuto12}, we find a 70~\micron\, $3~\sigma$ noise value of 3.07~Jy~beam$^{-1}$ and 2.03~Jy~beam$^{-1}$ toward SM1 and SM1N, respectively, corresponding to upper limits on the internal protostellar luminosity, $L_{int} \lesssim 0.9~L_\odot$ for SM1 and $L_{int} \lesssim 0.6~L_\odot$ for SM1N. We thus cannot rule out the presence of an embedded, low luminosity protostar toward either SM1 or SM1N based on the 70~\micron\, emission. 

Both FHSCs and young protostars are expected to drive outflows \citep[e.g., ][]{tomisaka02,machida05}. FHSCs are expected to drive slow outflows, with speeds of only a few \kms, in contrast to low-mass protostellar outflows with speeds of a $\sim 5-10$ \kms. In a search for outflows in Oph A, \citet{kamazaki03} find no evidence for outflow activity from either SM1 or SM1N in CO data.  The lack of outflow detection on large scales suggests that any outflow from either source must be dynamically young, in agreement with our estimates of the source ages. The FHSC candidates L1451-mm and Per-Bolo 58 drive outflows with a characteristic velocity of $< 3$~\kms\, \citep{pineda11,dunham11}. Over $\sim 10^3$~yr, a 3~\kms\, outflow would reach a distance of only 630~AU from the central source, whereas single-dish CO observations typically have spatial resolutions $\gtrsim 1800$~AU (e.g., CO 3-2 observed with the JCMT). A faster outflow with velocities $\sim 5-10$~\kms, typical of a Class 0 source, would drive material to $\sim 1000-2000$~AU in the same time frame. If either SM1 or SM1N contain FHSCs or an extremely young protostar, then any outflow will be best probed by future interferometric observations. 

\subsection{SM1: a protostellar object with an early disk}

Many Class 0 sources have been detected at radio wavelengths, where the emission is thought to be thermal bremmstrahlung (free-free) emission originating from the dense central base of ionized winds \citep{dzib13}. Radio surveys toward Oph A have not detected emission from SM1N, but show mixed results toward SM1. For example, \citet{dzib13} do not detect SM1 at 4.5~GHz and 7.5~GHz (6.7~cm and 4~cm, respectively) above 85~$\mu$Jy~beam$^{-1}$ (1\arcsec\, FWHM), but SM1 has been detected previously at 5~GHz (6~cm) with peak flux density $S_6 \sim 130-200~\mu$Jy~beam$^{-1}$ \citep[11\arcsec\, $\times$ 5\arcsec\, and 9\arcsec.5 $\times$ 5\arcsec\, FWHM, respectively; ][]{leous91,gagne04}. The \citeauthor{leous91} detection is $\sim 3$\arcsec\, offset from the ALMA peak location, whereas the \citeauthor{gagne04} detection is co-located with the ALMA source (see Figure \ref{fig:modImage}). For comparison, the 5~GHz flux densities found for the Class 0 source VLA 1623 in these studies are greater by factors of $\sim 1.5 - 2$. 

\citeauthor{gagne04} also find a faint, variable, and relatively hard X-ray source at the SM1 peak (shown in Figure \ref{fig:modImage}). No X-ray emission was seen toward VLA 1623, and searches for X-rays toward other known or possible Class 0 sources have had mixed results. For example, \citet{tsuboi01} detect hard X-ray emission toward several Class 0 sources in the Orion molecular cloud. Additional studies in different regions also find X-ray sources coincident with Class 0 objects \citep{hamaguchi05,getman06}. In contrast, \citet{giardino06} do not detect any X-ray emission toward known Class 0 sources in Serpens. In low-mass, pre-main-sequence stars, X-ray emission is likely caused by magnetic reconnection events on the stellar surface \citep{feigelson07}. Disk accretion processes may produce soft X-rays from accretion shocks, but the high extinction through Oph A limits the soft X-ray sensitivity toward SM1. 

Intriguingly, the SM1 core has a high degree of deuteration in H$_2$CO (Bergman et al. 2010) relative to SM1N, whereas the JCMT \hd\, observations at 11\arcsec\, resolution show a strong emission peak toward SM1N and little emission toward SM1. This seemingly discrepant result can be explained chemically: HDCO and D$_2$CO are known to form on dust grain surfaces (as intermediate products toward the formation of methanol, which can only form on dust surfaces) and partially desorb during the process of their formation in pre-stellar cores \citep[e.g., ][]{bacmann03,garrod07}. The bulk of deuterated formaldehyde, however, remains in ices, and is released back into the gas phase only nearby newly born protostars \citep[e.g., ][]{fuente05}. 

Together, the X-ray and radio detections strongly imply that SM1 is protostellar, containing an extremely young, deeply embedded protostellar object that is powering a partially ionized jet. In addition, the chemical abundances on single-dish scales are consistent with a source that has only recently begun to heat its immediate surroundings. These additional data thus support our characterization of the ALMA-detected continuum emission as a disk. The detection of a disk around SM1 is further evidence that disks can form rapidly around young protostars, coeval with or prior to the formation of a FHSC. 

Many, although not all, Class 0 protostars show compact dust continuum components that cannot be attributed to being simply the density profiles from their larger scale cores extrapolated to small scales, and are likely disks \citep[e.g., ][]{jorgensen04, jorgensen07, jorgensen09, girart09, maury14}. For these early protostellar stages, however, observational identification of disks has thus far largely relied on modelling the observed spectral energy distribution (SED), or unresolved structure in visibility data from interferometric observations. It is not clear whether all these early disks are rotationally supported; at least some Class 0 disks are dominated by infall rather than rotation \citep{brinch09}, as expected in early stages of core collapse when the young protostar is accreting most of its mass \citep[the formation of the circumstellar accretion region, or CAR; ][]{terebey84}. 

If the SM1 condensation is a circumstellar disk around an extremely young protostellar source, it is useful to compare the condensation mass ($M \lesssim 0.02$~M$_\odot$) with disks identified around other protostellar sources. Unless otherwise stated, we are comparing with studies that determine the disk mass from unresolved, small-scale components seen at long baselines in interferometric data (i.e., from the SMA or CARMA arrays). Disentangling the disk and envelope components thus add additional uncertainty to the disk masses beyond the temperature and opacity uncertainties described above. A simple comparison with the circumstellar disk around HL Tau \citep[following ][]{looney03} suggests the SM1 disk mass $M_{disk} \sim 0.005$~M$_\odot$ (where the HL Tau disk has a mass $M_{disk} = 0.05$~M$_\odot$ from a 2.7~mm flux density $S_\nu \sim 100$~mJy at a distance of 140~pc). A similar comparison for the Class 0 protostellar source L1157-mm gives a disk mass $M_{disk} = 0.004$~M$_\odot$  \citep{chiang12}.  \citet{enoch11} find a mean disk mass $M_{disk} \sim 0.2$~M$_\odot$ toward Class 0 and Class I sources in Serpens. Using similar assumptions about the disk temperature and dust opacity as used here, \citet{jorgensen09} find a median disk mass $M_{disk} = 0.089$~M$_\odot$ for a sample of ten nearby Class 0 sources, but note that incorporating a temperature gradient in the disk (as expected for a disk heated by a central protostar) would reduce the disk masses by a factor $\sim 2$. Recently, rotationally supported, Keplerian disks have been imaged toward two Class 0 sources \citep[L1527 IRS and VLA 1623; ][]{tobin13,murillo13a,murillo13b}, with radii $R \sim 70-180$~AU. Toward L1527 IRS, \citet{tobin13} find a disk mass $M \sim 0.007$~M$_\odot$. The masses of these early disks are in good agreement with our mass estimate in the SM1 condensation, despite most having significantly higher protostellar luminosities than our upper limit toward SM1.  

\subsection{SM1N: a possible FHSC?}

In contrast to SM1, SM1N shows no evidence for an embedded source in X-ray, infrared, or radio emission. Figure \ref{fig:modImage} shows that after $uv$-subtraction of the best-fit Gaussian model, the residual image shows no additional small-scale structure, and we detect no point source above our noise limits when imaging the longest baselines only.  Toward SM1N, the ALMA-detected continuum structure is significantly smaller than the scales over which starless cores are modelled to have relatively flat density profiles \citep[a few thousand AU; ][]{ward-thompson94,kirk05}. The ALMA data thus reveal the accumulation of mass on small scales, again at a time just prior to or after the formation of a FHSC or protostar. 

The SM1N condensation is significantly larger than predicted FHSC radii of only a few AU \citep[e.g., ][]{masunaga98,tomida10,tomida13}. Rotating FHSCs may form early circumstellar disks with radii up to $\sim 100$~AU, but these are only seen in simulations of the fastest rotating cases \citep[e.g.,][]{bate98,saigo06,saigo08,machida11,bate11}. The SM1N condensation is similar in extent, however, to that expected from theoretical predictions for a pseudodisk, a flattened, disequilibrium structure expected to form around FHSCs and young protostars as material is accreting onto the central source in the presence of a magnetic field \citep{galli_shu1,galli_shu2}. \citet{commercon12} simulate ALMA observations of FHSCs from radiation-magneto-hydrodynamic models of collapsing dense cores at 150~pc, and find detectable accretion disks with sizes similar to the SM1N condensation for cores with initial moderate and strong magnetic fields. While the predicted flux density in the simulated extended structure is similar to what we find toward SM1N, the predicted peak flux from the central FHSC is $\sim 10 $ times brighter than the observed peak flux toward SM1N. \citet{furuya12} perform three-dimensional radiative transfer modelling of the collapse from core to FHSC. Evaluating the chemistry post-process, the authors find that heating allows CO to sublimate at radii of $\sim 40 - 100$~AU from the FHSC, which would reduce the \hd\, abundance due to both the gas temperature and the increased CO abundance. The ALMA detection of \hd\, toward SM1N is therefore consistent with the presence of an embedded FHSC within SM1N.

The elongation of the SM1N condensation follows that seen on larger scales in \dia\, emission \citep{difran04}, matching the filamentary shape of the northern tip of Oph A. If SM1N is truly starless, this observed alignment supports the prediction that the collapse of filamentary structures should create triaxial structure on small ($\lesssim 5000$~AU) scales \citep{lee12}, eventually forming filamentary envelopes similar to those seen around some Class 0 protostars \citep{tobin10}. Similar agreement between core and filament elongation over larger scales has been seen toward the starless Oph AN6 core \citep{difran04,bourke12} and Barnard 5 \citep{pineda11a}. We note, however, that the \hd\, line widths observed by ALMA suggest any non-thermal (i.e., infall) motions are likely subsonic. More sensitive observations of undepleted, spatially resolved molecular line tracers are needed to probe the infall and velocity gradients expected in this filament accretion scenario. 

\subsection{Substructure in clustered cores}

Both the SM1 and SM1N condensations are smaller than the Jeans lengths derived from their respective mean densities. We therefore do not expect thermal fragmentation to drive further substructure formation. If the SM1 condensation is a disk, its gravitational stability will depend, for example, on whether or not it is in Keplerian rotation, and the mass of the disk relative to the central source \citep[e.g., ][]{toomre64}. Additional observations are needed to determine both of these parameters. 

Toward SM1, \citet{nakamura12} detect several small-scale condensations with the SMA through 342~GHz (880~\micron) observations at an angular resolution of $2.\arcsec7 \times 1.\arcsec3$. The brightest SMA source (a1) is coincident with the ALMA-detected condensation, and has a peak flux density that is consistent with the total ALMA flux. We do not detect a second source (a2), although it should be visible within the ALMA field of view, with a peak flux density that is significantly greater than our rms noise ($\sim 0.1$~Jy~beam$^{-1}$). The $uv$-data in Figure \ref{fig:modUV} shows that SM1 contains additional structure at larger spatial scales in addition to the Gaussian model, but the `ringing' feature in $uv$-space is consistent with an extended feature offset in position from the primary source, rather than a secondary Gaussian feature at lower brightness. The most likely source of this ringing is the extended Oph A ridge, and if we image the residual data (after subtraction of the SM1 Gaussian model, shown in Figure \ref{fig:modImage}) with natural weighting, we indeed see a low-brightness, extended feature to the west of SM1 and elongated along a north-south axis, consistent with the Oph A ridge at the location of SM1. 

The observations described here are some of the first to probe the substructure in starless or very early protostellar cores at $\sim 50$~AU resolution. Lower resolution studies have similarly detected little substructure in starless cores \citep{schnee10,schnee12} but likely lack the sensitivity and resolution to discern fragmentation in these sources \citep{offner12}. Here, we show that the SM1 and SM1N cores are unlikely to fragment into additional structures to form, e.g., binary or multiple stellar systems. It is not necessarily true, however, that SM1 and SM1N could not eventually end up in a multiple system.  \citet{difran04} find that to first order, Oph A is in gravitational equilibrium. They also find that the core-to-core velocity dispersion along the line of sight, $\sigma_r$, within Oph A is only $0.12$~\kms. Assuming the three-dimensional velocity dispersion $\sigma_{3D}^2 = 3 \sigma_r^2$, and given a projected distance between SM1 and SM1N of 35\arcsec.5 ($\sim 4300$~AU at a distance of 120~pc; using the peak continuum emission locations determined here), we find an interaction timescale of only $t_{int} = 1 \times 10^5$~yr. This is significantly shorter than estimates of the prestellar core timescale \citep[0.5~Myr; ][]{enoch08}, and the embedded core lifetime \citep[0.5~Myr; ][]{evans09}, indicating that the two objects could interact before the dispersal of the Oph A filament. 

\subsection{\hd\, in cores at $\sim 50 - 100$~AU resolution}

Figure \ref{fig:hdspec} shows that the \hd\, emission detected toward SM1N is unexpectedly offset to the east from the central continuum peak. Here, we investigate several potential causes of this offset. 

Enhanced abundances of \hd\, in the gas phase within dense cores require low gas temperatures. Since the \amm-derived gas temperature on $\sim 1000$~AU scales is 16~K, an increasing temperature gradient to $T \gtrsim 20$~K toward the core centre could reduce the production of \hd\, through two mechanisms. First, the reaction to form \hd\, is exothermic by only a small amount \citep[$\Delta E = 232$~K; ][]{millar89}, and at higher temperatures it can proceed in both directions, producing no overabundance of \hd. Second, at higher temperatures CO will evaporate from dust grains \citep{bisschop06} and react with \hd, lowering the \hd\, abundance. In this scenario, SM1N contains an unseen low luminosity central source heating the gas and dust.  

\citet{caselli08} show that a dust grain size distribution shifted to smaller grains and higher cosmic ray ionization rate will both decrease the observed deuterium fractionation at low temperatures. Since grains are expected and observed to coagulate and grow at high densities, we do not expect the grain size distribution to play a role in the observed \hd\, distribution toward SM1N. Furthermore, there is no evidence of a variation in the ionization rate toward Oph A. We note, however, that immediately east of Oph A there is a strong, non-thermal radio and X-ray source, Oph S1, a close T-Tauri binary with a magnetic B4 primary \citep[J162634.2-242328; ][]{gagne04}. Given the X-ray luminosity detected by \citeauthor{gagne04}, we follow \citet{maloney96} to determine the X-ray ionization rate at the location of SM1, and find that it is less than the typical cosmic ray ionization rate assumed for molecular clouds. 

Recent chemistry models that include grain surface reactions in dense cores show that deuterium can become incorporated into HDO on grain surfaces at densities $n \gtrsim 10^6$~\cc\, and temperatures $T \lesssim 17$~K, significantly reducing the abundance of HD (and consequently \hd) in the gas phase \citep[][Lee \& Bergin 2014, submitted]{sipila13}. This effect is strongest at high densities and low temperatures ($n \gtrsim 10^6$~\cc\, at $T \sim 10-15$~K), resulting in a decreasing \hd\, abundance gradient toward small radii in a starless core. Given the physical conditions we derive for SM1N (a mean density of $n \sim 10^7$~\cc\, and a gas temperature $T \sim 16$~K), the \hd\, abundances we find are consistent with model predictions. For a model core with slightly lower average density than we find for the SM1N condensation (model $<n> \sim 8 \times 10^5$~\cc, with $n \sim 2 \times 10^6$~\cc\, at radii $\lesssim 500$~AU), \citeauthor{sipila13} predict an \hd\, abundance decrease of an order of magnitude, to $X(\mbox{\hd}) < 10^{-10}$, at radii $\lesssim 500$~AU.  Assuming the density profile increases and the temperature profile decreases toward the core centre, Lee \& Bergin predict a drop in the gas-phase \hd\, abundance (including both ortho- and para-\hd) to values $\lesssim 10^{-11}$. These updated chemistry models can therefore also explain the lower \hd\, abundance at the SM1N continuum peak, but do not predict directly the observed asymmetric distribution. 

The observed small-scale \hd\, emission, along with the lack of other indicators of a core's protostellar nature, further support our conclusion that SM1N is either an evolved, starless core, or contains a very young protostellar (possibly FHSC) source, where the ALMA-detected condensation reveals the pseudodisk around the YSO or FHSC. 

\section{Summary}
\label{sec:summary}

We have presented ALMA observations of the Ophiuchus A SM1 and SM1N cores in submillimeter dust continuum emission, and ALMA and JCMT observations of \hd\, $1_{10}-1_{11}$ emission. We detect small-scale continuum condensations within both SM1 and SM1N. Both condensations have masses $M \lesssim 0.02$~M$_\odot$ ($\sim 20$~M$_{Jup}$) based on the continuum emission and typical dust temperature and dust emissivity values in prestellar cores or protostellar disks. We show that the SM1 condensation is consistent with a nearly-symmetric Gaussian source with a width of only $37$~AU, while the SM1N condensation is elongated, and extends 500~AU along its major axis. No evidence for substructure is seen in either source. A Jeans analysis indicates they are unlikely to fragment, suggesting both cores will form single stars. We propose that SM1 is protostellar, and that the condensation detected by ALMA is a warm ($T \sim 30-50$~K) accretion disk or pseudodisk. The less concentrated emission of SM1N suggests that it is still starless, but we cannot rule out the presence of a low-luminosity source. 

We furthermore show the first interferometric \hd\, detection toward a star-forming region. \hd\, is only detected toward SM1N, but is offset from the continuum peak. This offset may be due to either heating from an undetected, young, low luminosity protostellar source or first hydrostatic core, or HD (and consequently \hd) depletion in the cold centre of the condensation. These data reveal observationally the earliest stages of the formation of circumstellar accretion regions, and agree with theoretical predictions that disk formation can occur very early in the star formation process, coeval with or just after the formation of a first hydrostatic core or protostar. The identification of SM1 (and possibly SM1N) as protostellar indicates that infrared surveys may not produce complete catalogs of protostellar cores in high extinction environments, such as Ophiuchus. 

\acknowledgments

This paper makes use of the following ALMA data: ADS/JAO.ALMA\#2011.0.00396.S. ALMA is a partnership of ESO (representing its member states), NSF (USA) and NINS (Japan), together with NRC (Canada) and NSC and ASIAA (Taiwan), in cooperation with the Republic of Chile. The Joint ALMA Observatory is operated by ESO, AUI/NRAO and NAOJ. The authors thank B. Matthews for performing the initial reduction of the ALMA data. The National Radio Astronomy Observatory is a facility of the National Science Foundation operated under cooperative agreement by Associated Universities, Inc. The James Clerk Maxwell Telescope is operated by the Joint Astronomy Centre on behalf of the Science and Technology Facilities Council of the United Kingdom, the National Research Council of Canada, and (until 31 March 2013) the Netherlands Organisation for Scientific Research. The Dunlap Institute is funded through an endowment established by the David Dunlap family and the University of Toronto. The research of JKJ is supported by a Junior Group Leader Fellowship from the Lundbeck Foundation. JEP is supported by the Swiss National Science Foundation, project number CRSII2\_141880.

\bibliographystyle{apj}
\bibliography{biblio}

\end{document}